\documentclass[secnumarabic,amssymb, nobibnotes, aps, prd]{revtex4}

\usepackage{amssymb}
\usepackage{amsmath}
\usepackage{bm}
\usepackage{graphicx,color}
\usepackage{color}

\begin{document}

\title{Decoherence of scalar cosmological perturbations}%
\author{Mariano Franco}
\email[]{marianoandresfranco@gmail.com}
\author{Esteban Calzetta}
\email[]{calzetta@df.uba.ar}
\affiliation{Departamento de Fisica, FCEN-UBA and IFIBA-CONICET, Pabellon I, 1428, Buenos Aires, Argentina}
\begin{abstract}
In this paper we analyze the possibility of detecting nontrivial quantum phenomena in observations of the temperature anisotropy of the cosmic background radiation (CBR), for example, if the Universe could be found in a coherent superposition of two states corresponding to different CBR temperature self-correlations. Such observations are sensitive to scalar primordial fluctuations but insensitive to tensor fluctuations, which are therefore converted into an environment for the former. Even for a free inflaton field minimally coupled to gravity, scalar-tensor interactions induce enough decoherence among histories of the scalar fluctuations as to render them classical under any realistic probe of their amplitudes.
\end{abstract}
\maketitle

\section{Introduction}
According to the inflationary paradigm \cite{Wei,Bass}, not only primordial cosmological fluctuations are quantum in origin, but they are also created in a highly non-classical state \cite{CamPar06,CamPar08a,CamPar08b,MazHey09}. This raises the tantalizing possibility of uncovering nontrivial quantum behavior through cosmological observations. However, no known cosmological probe would reveal the actual quantum state of primordial fluctuations: all known methods of observation focus on a restricted set of properties of those fluctuations, thus leaving a remainder which must be traced over. Therefore, to discuss nontrivial quantum behavior we have to consider not only the quantum features of the cosmological fluctuations, but also the loss of quantum coherence induced by the partial description appropriate to the observation in question.

In this paper we take as example the case of observations of the amplitudes of the temperature anisotropy of the cosmic background radiation (CBR). An example of such an observation is the determination of the temperature self-correlation. The temperature fluctuations are determined by the scalar cosmological fluctuations. Unlike the case when CBR polarization is being observed,   tensor perturbations affect the result only through their action on the scalar ones. Therefore in the observation of CBR temperature fluctuation amplitudes, we must regard tensor fluctuations as an environment coupled to the system of interest, namely the scalar fluctuations.

The coupling between the system and its environment induces decoherence in the former \cite{JZKG,Schl}. 
Adopting the Hartle- Gell-Mann consistent histories approach to quantum mechanics \cite{Gell-Har,Har}, we ask whether it is possible to observe the coherence between different histories of the scalar fluctuations, after tracing over the tensor fluctuations. We shall only consider the coupling between these fluctuations demanded by general relativity. We represent all matter fields by a single free scalar inflaton field, minimally coupled to gravity. After identifying the relevant gauge invariant variables and imposing the Newtonian gauge conditions (see below), the momentum constraints of general relativity relate the inflaton field to the single scalar degree of freedom in the metric, so there is only one gauge invariant scalar degree of freedom in the theory. This scalar field is coupled to the graviton field, which after making the graviton polarization explicit may also be described by two scalar fields. We disregard vectorial perturbations. 

Our conclusion is that the decoherence induced by tensor perturbations is strong enough to erase any traces of quantum behavior in the scalar fluctuations, given any realistic observational scenario by today's standards. To this extent, our findings are consistent with other treatments of the issue in the literature, based on different system-environment splits, or else on averaging over the decaying mode of the cosmological fluctuations \cite{GP,CalHu95,CalGon97,
KP,KLPS,KPS,Lom-Lop,ProRig07,Bur,Mar}.

Within the Gell-Mann and Hartle formalism one has the freedom to take any pair of histories to compute the decoherence functional. We choose these histories to answer if quantum effects in the CBR spectrum can be detected. To link scalar perturbations with temperature fluctuations in CBR we use the Sachs-Wolfe effect \cite{Dod,Lin,Wei}. But this is only valid in the recombination era, to link the amplitude of scalar perturbations with the inflationary period we use Bardeen's conservation law \cite{Bard,MFB} and later we can replace the scalar perturbations in inflation by the fluctuations in CBR. Then, we finally compute the decoherence induced by the gravitons on two histories of the CBR fluctuations and we see how much this histories can differ of the observed Harrison-Zel'dovich spectrum in order to see the quantum effects in the CBR fluctuations. 

This paper is organized as follows. In Section II we compute the interaction between the scalars and tensors modes which is necessary to calculate the decoherence functional (the details of calculations are left to the Appendix B.3). Section III is devoted to compute the decoherence functional (details of the calculation are given in Appendix C), and we compare our results with several works mentioned above. Finally, Section IV contains our conclusions. In Appendix A we give a brief summary of the Hartle- Gell-Mann approach, mainly to fix our notation. For the same reason, a brief summary of inflation, gauge invariant cosmological perturbations and their link to CBR temperature is given in Appendix B. We use units such that $\hbar=c=k_{B}=1$ and $G=m^{-2}_{pl}$.

\section{Inflation and cosmological perturbations}

The aim of this work is to calculate the decoherence suffered by the scalar perturbations due to its interaction with the tensor perturbations in the inflationary stage of the Universe. For such calculation it is necessary to find the interaction between the perturbations. In this section we calculate the interaction between scalar and tensor modes using the ADM formulation of General Relativity \cite{MTW}. Later we compute the free action of the tensor perturbations and the Hadamard propagator \cite{CalHu} associated to them. We will use the \textit{Newtonian gauge} to the cosmological perturbations \cite{MFB} (see Appendix B.2 to details). 

\subsection{Scalar-Tensor interaction}

The calculation of interaction between scalars and tensor perturbations was performed using the ADM formulation of general relativity \cite{MTW}. The total action in this formulation is the Einstein-Hilbert action for gravity together with the Klein-Gordon action for the inflaton. Our notation is explained in full in Appendix B.

Using (\ref{nodiag2}) in equation (\ref{nodiag1}) we obtain one equation that links the scalar modes \cite{MFB}

\begin{equation}
 \phi'+\mathcal{H}\phi=4\pi m^{-2}_{pl} \varphi'_{0}\delta\varphi
\end{equation} 

Then, in the Newtonian gauge a single degree of freedom remains a scalar. Now all we have to do is to find the coupling between this perturbation and tensor modes.

We compute the interaction between one graviton and two scalars \cite{Mal, W}. A complete treatment is done in  Appendix B.3. 

Combining (\ref{inter1}), (\ref{inter2}), (\ref{inter3}) and (\ref{inter4}) we obtain 

\begin{equation}
S_{int}=\int d^4x J_{ij} h_{ij}
\label{Sint}
\end{equation}

\begin{widetext}
\begin{equation}
J_{ij}=\frac{1}{2}m^2_{pl} a^2(\eta) \left[ \left(
1+16m^2_{pl}\tilde{H}^2\varphi'^{-2}_0\right) \phi_{,i}\phi_{,j}+16m^2_{pl}\varphi'^{-2}_0\phi'_{,i}\phi'_{,j}+32
m^2_{pl}\tilde{H}\varphi'^{-2}_0\phi_{,i}\phi'_{,j}\right]
\end{equation}
\end{widetext}	
This is the coupling current that is used to calculate the decoherence induced on the scalar tensor modes. But to do this is also necessary to calculate the Hadamard propagator: the equation (\ref{fd}) shows that the decoherence functional is written in terms of the Hadamard propagator and to compute it requires first to find the free action for the tensor perturbations.

To second order in $h_{ij}$, the free action of the gravitons is the Klein-Gordon action for tensors $h_{ij}$  

\begin{equation}
 S_{free}=\frac{m^2_{pl}}{4}\int d^4x a^2(\eta)\left[
h'_{ij}h'_{ij}-h_{ij,k}h_{ij,k}\right] 
\end{equation} 

The free dynamics of the gravitons is described in terms of their physical degrees of freedom \cite{CalHu95}

\begin{equation}
 h_{ij}(\eta,x)=\frac{1}{m_{pl}}\int d^3y
[G^+_{ij}(x-y)h^+(\eta,y)+(+\leftrightarrow \times)]
\end{equation}  

where $+$ and $\times$ are the graviton polarizations and

\begin{equation}
 G^+_{ij}(x-y)=\int \frac{d^3k}{(2\pi)^3}e^{ik(x-y)}A^{+k}_{ij}
\end{equation} 

The matrix $A_{ij}$ verifies

\begin{equation}
 A^{+k}_{ii}=k_iA^{+k}_{ij}=A^{\times k}_{ii}=k_i A^{\times k}_{ij}=0 \label{tepol}
\end{equation} 

and $h(\eta, y)$ obeys

\begin{equation}
 h''+2\frac{a'}{a}h-\nabla^2h=0
\end{equation} 

where the scalar field $h(\eta,y)$ is in the usual Bunch-Davis vacuum state.

The Hadamard propagator is defined as

\begin{equation}
 G_1(\eta,y;\eta',y')=\langle h(\eta,y)h^*(\eta',y')+h^*(\eta,y)h(\eta',y') \rangle
\end{equation} 

It results

\begin{widetext}
\begin{equation}
G_1(\eta,y;\eta',y')= \int\frac{d^3k}{(2\pi)^3} \frac{1}{2k}\left\lbrace 
(1+\frac{1}{k^2\eta\eta'})\cos[k(\eta-\eta')+k(y-y')]\frac{1}{k}(\frac{1}{\eta}-\frac{1}{\eta'})\sin[k(\eta-\eta')+k(y-y')]\right\rbrace 
\end{equation}  
\end{widetext}

In the next Section we compute the decoherence functional using the results developed in this Section together with the results of Appendix B.4. In this Appendix we relate scalar cosmological perturbations to CBR temperature fluctuations. This allows us to rewrite the decoherence functional in terms of an observable.

\section{Decoherence Functional}

The Hartle Gell-Mann formalism lets us choose between the histories involved in the decoherence functional. In this work we wish to choose histories representing different outcomes regarding the CBR temperature. Since nonlinear effects are small, the CBR temperature is determined by the scalar perturbations, and these evolve as a nearly free field. Therefore we assume histories where the single gauge invariant scalar perturbation $\phi(\eta)$  allowed in the Newtonian gauge evolves as a free perturbation, as described in Appendix B, while tensor perturbations are totally unspecified. 

Such a history is determined by the amplitudes $\hat{a}_k$ and $\hat{a}_k^{\dagger}$ in eq. (\ref{freescalar}), which we determine as follows. Observe that we envisage a situation where the amplitude of the temperature anisotropies may be detected, but not the phases of individual modes. Therefore we write the amplitudes $\hat{a}_k$ and $\hat{a}_k^{\dagger}$ in terms of absolute value and phase as \cite{CalHu,Sta}

\begin{equation}
\hat{a}_k\rightarrow \sqrt{L^3}f_\textbf{k} e^{i\theta_k} \label{stoc}
\end{equation}

where $f_k$ is a stochastic function that depends on the $k$ mode and $\theta_k$ is a phase that depends on the mode and takes random values between $0$ and $2\pi$. By assuming independent phases for different modes within a given resolution (see below, eq. (\ref{RPA})) this ansatz reproduces a translation invariant correlation function for the temperature fluctuations. The $L^3$ factor is the co-moving volume where the stochastic functions are defined and it is taken as the volume where the inflaton field is homogeneous at the beginning of inflation. This volume can be estimated using the slow-roll and vacuum dominance conditions (we have used that $a_i\equiv 1$, with this convention the co-moving and physical volume are equals at the beginning of inflation),

\begin{equation*}
 \left(\nabla \varphi \right)^2<<V(\varphi)\Rightarrow \varphi^2/L^2<<V(\varphi)
\end{equation*} 

and replacing the inflaton field by (\ref{inflaton}) and the potential energy by (\ref{Hminflaton}) we obtain

\begin{equation}
L>>1/(\sqrt{\epsilon}H) \label{vacumm}
\end{equation}  

Therefore, introducing (\ref{stoc}) in the scalar perturbation and expanding in Fourier modes equation (\ref{SW}) we obtain for each mode

\begin{equation}
\phi_k(\eta_i)e^{ik\eta_e}f_\textbf{k}=-\frac{\dot{\varphi}^2}{V(\varphi)}\frac{\delta T_k}{T_0} \label{CBR}
\end{equation}

With this last equation we can relate the scalar perturbation modes during inflation with the CBR anisotropies, which is an observable magnitude.

We start the decoherence functional computation by writing it in terms of the coupling current, the Hadamard propagator and the polarization tensors of the gravitons as

\begin{widetext}
\begin{equation}
Im(S_{FI})=\int d^4x\;d^4x'\int  d^3y\;d^3y' \int\frac{d^3k}{(2\pi)^3} \frac{d^3k'}{(2\pi)^3}\frac{d^3k''}{(2\pi)^3}e^{ik(x-y)}e^{-ik'(x'-y')}
\frac{1}{m^2_{pl}} A^k_{ij}A^{k'}_{lm} C_{ij}(x)C_{lm}(x')
 G_{1k''}(\eta,y;\eta',y')
\end{equation}  
\end{widetext}

where $C_{ij}(x)= J^1_{ij}(\phi^1(x))-J^2_{ij} (\phi^2(x))$ is the difference between the two histories involved in the decoherence functional. 

We develop the interaction between the scalar and tensor modes into Fourier modes using that when $\phi_k$ is within the horizon it verifies $\phi'_k=ik\phi_k$ (see Appendix B.4). Each coupling current has a series expansion as follows

\begin{equation*}
J_{ij}(x)=\frac{m^2_{pl}}{2}a^2(\eta)\int \frac{d^3q}{(2\pi)^3}\frac{d^3p}{(2\pi)^3} q_ip_j \hat{A}_{qp}
\end{equation*} 
\begin{equation}
\left[\left(1+16m^2_{pl}\varphi'^{-2}_0\mathcal{H}\right)-16m^2_{pl}\varphi'^{-2}_0qp+ i 32m^2_{pl}\varphi'^{-2}_0p\mathcal{H}\right] 
\end{equation} 

with $\hat{A}_{qp}=\hat{\phi}_q(\textbf{x})\hat{\phi}_p(\textbf{x})$, and the scalar perturbation has a development as

\begin{equation}
\hat{\phi}(\textbf{x})=\phi^*_q(\eta)e^{i\textbf{qx}}\hat{a}^-_q+\phi_q(\eta)e^{-i\textbf{qx}}\hat{a}^+_q
\end{equation}

Replacing this development in the decoherence functional and taking the product $C_{ij}(x)C_{lm}(x')$ we obtain,

\begin{widetext}
\begin{equation*}
Im(S_{IF})= \int  d^4xd^4x'd^3yd^3y' \frac{d^3k}{(2\pi)^3} \frac{d^3k'}{(2\pi)^3} \frac{d^3k''}{(2\pi)^3}\frac{d^3q}{(2\pi)^3}\frac{d^3p}{(2\pi)^3} \frac{d^3q'}{(2\pi)^3}\frac{d^3p'}{(2\pi)^3}
\end{equation*}
\begin{equation*}
A^k_{ij}A^{k'}_{lm}q_ip_jq'_lp'_m e^{ik(x-y)}e^{ik'(x'-y')} G_{1k''}(y,\eta;y',\eta')m^4_{pl}a^2(\eta)a^2(\eta')
\end{equation*} 
\begin{equation*}
\left[1+16m^2_{pl}\varphi'^{-2}_0\mathcal{H}^2-16m^2_{pl}\varphi'^{-2}_0qp+i32m^2_{pl}\varphi'^{-2}_0p\mathcal{H}\right](\eta)
\left[1+16m^2_{pl}\varphi'^{-2}_0\mathcal{H}^2-16m^2_{pl}\varphi'^{-2}_0q'p'+i32m^2_{pl}\varphi'^{-2}_0p'\mathcal{H}\right](\eta')
\end{equation*} 
\begin{equation}
\left\lbrace \phi^*_q(\eta) \phi^*_p(\eta) \phi_{q'}(\eta') \phi_{p'}(\eta') e^{i(\textbf{q}+\textbf{p})\textbf{x}}e^{-i(\textbf{q}'+\textbf{p}')\textbf{x}'}\right.
\left.(\hat{a}^{-1}_q\hat{a}^{-1}_p-\hat{a}^{-2}_q\hat{a}^{-2}_p)(\hat{a}^{+1}_{q'}\hat{a}^{+1}_{p'}-\hat{a}^{+2}_{q'}\hat{a}^{+2}_{p'})+\text{other comb}\right\rbrace 
\end{equation} 
\end{widetext}

where \textit{other comb} refers to all possible combinations of the product of the four $\phi$ alternating conjugated and conveniently changing the sign of the exponential in the creation and destruction operators. Using all the possible combinations sixteen terms are obtained. From here on we develop one of those terms only, the computation of the other terms is fully analogous.

Now it is convenient to integrate in $d^3y$ and $d^3y'$, because only the Hadamard propagator (via sines and cosines of $ky$ and $k'y'$) and the exponentials in the tensors polarization depend on these variables. For this it is suitable to write the propagator as

\begin{equation*}
G_{1k''}=\frac{1}{m^2_{pl}}\left[\alpha_{k''}\left(e^{i\textbf{k}''(\textbf{y}-\textbf{y}')}+e^{-i\textbf{k}''(\textbf{y}-\textbf{y}')}\right)+\right.
\end{equation*}
\begin{equation}
\left.\beta_{k''}\left(e^{i\textbf{k}''(\textbf{y}-\textbf{y}')}-e^{-i\textbf{k}''(\textbf{y}-\textbf{y}')}\right)\right] \label{alfa}
\end{equation}  

where $\alpha_k=k^{-1}(1+k^{-2}\eta^{-1}\eta'^{-1})$ and $\beta_k=k^{-2}(\eta^{-1}-\eta'^{-1})$.

Integrating in $d^3y$ y $d^3y'$ arise Dirac's deltas combining the modes $\textbf{k}$, $\textbf{k}'$ and $\textbf{k}''$ (see (\ref{fd1})). After using this deltas to reduce the number of modes (only $\textbf{k}''$ remains and we recall it as $\textbf{k}$ to simplify) it is convenient to integrate in $d^3x$ y $d^3x'$ using the Fourier expansion of each $\phi$. This is shown in (\ref{fd2}). With this integration appear a second term with the same combination of modes $\phi_k$ and $\phi^*_k$ and is due to the emergence of different deltas to integrate in the mode that developed the Hadamard propagator.

Next we approximate the phase factors by their expectation value  according to the rule

\begin{equation}
e^{i\left(\theta_q-\theta_{q'}\right)}\approx\left\langle e^{i\left(\theta_q-\theta_{q'}\right)}\right\rangle\approx 1 \Leftrightarrow \arrowvert \textbf{q}-\textbf{q}'\arrowvert << \delta q'
\label{RPA} 
\end{equation} 

and $0$ otherwise. Here $\delta q'$ denotes our resolution in momentum space. We assume $\delta q' \sim \sqrt{\epsilon}H\sim L^{-1}$, where $L$ has been defined in eq. (\ref{vacumm}).

After this we rewrite the decoherence functional in terms of fluctuations in the temperature of the CBR through (\ref{CBR}). This is done in equation (\ref{fd4}).

The next step is to consider the histories involved in the decoherence functional. These histories are parameterized by CMB fluctuations. We consider histories which lead to identical CBR temperatures except in a small window in mode space. One of these histories leads to the usual Harrison-Zel'dovich spectrum ($\delta_2 T( \textbf{k}) \sim \arrowvert \textbf{k}\arrowvert^{-3/2}$, see eq. (\ref{THZ}) below) and the another history ($\delta_1 T(q)$) remains unspecified. The window is centered at $\textbf{q}_0$, where this mode left the horizon at some moment of inflation, $\vert \textbf{q}_0 \vert \sim a_q H$, where $a_q$ is the scale factor when $\textbf{q}_0$ crosses the horizon. This window has a width $\delta q_0 \sim \Delta l/r_{LS}$, being $\Delta l$ the range of multipoles included in the window and $r_{LS}$ the co-moving distance to the last scattering surface. This distance is $r_{LS}=\int^{t_0}_{t_rec} \; a^{-1}(t)dt$, where $t_0$ is the present time, $t_{rec}$ is the recombination time and the scale factor is 

\begin{equation}
a(t)=e^{60}\frac{T_{reh}}{T_0}\left(\frac{t}{t_0}\right)^{2/3} \label{scalefactor}
\end{equation} 

where $T_{reh}\sim 10^{16}GeV$ is the reheating temperature and $T_0\sim 10^{-4}ev$ is the present temperature. This renormalization is to be consistent with the previous definitions, $a_i\equiv 1$ and $a_f \sim e^{60}$. The distance to the last scattering surface results $r_{LS}\sim e^{60} 10^{-29}t_0$, with $t_0\sim 10^{17}s\sim 10^{42}GeV^{-1}$.

The computation of the decoherence functional for these histories is long but straightforward; we give details in Appendix C. To write down the final result, let us introduce the scale-invariant (Harrison-Zel'dovich) temperature spectrum

\begin{equation}
\delta T_{HZ}\left[q\right]=T_0\frac{N_{\phi}}{q^{3/2}}
\label{THZ}
\end{equation}
where $N_{\phi}\sim 10^{-5}$. The decoherence functional is

\begin{widetext}
\begin{equation}
Im(S_{IF}) \approx  \frac{m^2_{pl}\epsilon^4}{H^4}N^2_\phi L^3 (\delta q_0)^3 \frac{I(q_0)}{q^3_0}\;\left[\frac{\delta T_1-\delta T_2}{\delta T_{HZ}}\right]^2(q_0)
\end{equation}
\end{widetext}

$I(q_0)$ is given in eq. (\ref{fd7}). A numerical analysis shows that $I(q_0)/q^3_0$ is essentially constant; it can be substituted by a numerical factor of the order of $10^{78}$ (omitting the $\frac{H^2}{\epsilon^2}$ factor that comes from the parameterization). We use this factor to complete the calculation of the decoherence functional.

Finally, we have that

\begin{equation}
Im(S_{IF}) \approx 10^{78}\;\frac{m^2_{pl}\epsilon^2}{H^2}N^2_\phi \; L^3 \;(\delta q_0)^3 \left[\frac{\delta T_1-\delta T_2}{\delta T_{HZ}}\right]^2 (q_0)
\end{equation}

To estimate the numerical factor that involves $m^2_{pl}\epsilon^2H^{-2}$ we take $m_{pl}\sim 10^{19}GeV$, $H\sim10^{13}GeV$ and $\epsilon\approx 10^{-2}$. Furthermore, we replace the factors  $\delta q_0 \sim \Delta l/r_{LS} $ and $L\sim 1/(\sqrt{\epsilon}H)$. 

Then, the decoherence functional results

\begin{equation}
Im(S_{IF}) \approx 10^{79}\Delta l^3 \left[\frac{\delta T_1-\delta T_2}{\delta T_{HZ}}\right]^2(q_0) \label{result}
\end{equation} 

Analyzing those results we see that decoherence is most effective for modes leaving the horizon towards the end of inflation. 

\section{Conclusions}
In this paper we have computed the decoherence functional between two histories of the Universe where scalar primordial fluctuations evolve in a prescribed way while tensor fluctuations are regarded as an environment. This decoherence functional is relevant to the question of whether it is possible to detect nontrivial quantum behavior in observations of the CBR temperature alone (that is, blind to CBR polarization). Our result implies that such detection is unrealistic by today standards. Because of the well known triangle inequality, we expect the same would obtain if the scalar fluctuations were regarded as an environment for the tensor ones.

This finding is consistent with earlier analysis of decoherence of cosmological fluctuations \cite{GP,CalHu95,CalGon97,
KP,KLPS,KPS,Lom-Lop,ProRig07,Bur,Mar}. We hold this paper is an advance with respect to those earlier analysis because our system-environment split is related to the features of a realistic observational scheme, rather than just being assumed. Moreover, we make no ad-hoc assumptions regarding the model, since the only coupling we are considered is demanded by general relativity. The present work is probably closest to \cite{CalHu95}, but goes beyond it in that the proper gauge invariant degree of freedom is identified, rather than just the inflaton field.
 
Finding tangible proof of the quantum nature of our Universe is one of the most fascinating challenges faced by Cosmology today. We believe our result should not be read in a negative way but rather in a positive one, as pointing to the direction in which a successful scheme could be found. We are continuing our research with this goal in mind.

This work is supported by the University of Buenos Aires, CONICET and ANPCyT. We acknowledge valuable discussions with D. Lopez Nacir, F. Lombardo and F. D. Mazzitelli.

\appendix

\section{Decoherence Functional}
In this section we give a quantitative discussion of decoherence. To calculate the loss of coherence induced on the scalar tensor modes which are in the FRW metric we use the decoherence functional developed for Gell-Mann and Hartle \cite{Gell-Har, Har}. We give a brief description of closed quantum systems including the decoherence term that is related to the classical sum rule of probabilities to different histories of a closed quantum system.

In the consistent histories description there is a subset of configuration space variables that are distinguished ($\psi$, system) while another subset is ignored ($\xi$, environment). An individual coarse-grained history is described by the path $\psi^\alpha(t)$ along with all possible paths $\xi^\alpha(t)$.

When the probability of each history can be assigned individually, the system behaves like a classical one and we say it has decohered. This means that the quantum interference between this set of histories is negligible and the probability of reaching the same final state through two different stories is the sum of probability of each history. The interest in finding histories that have undergone decoherence lies in the fact that these histories will be the ones that describe the classical domains.

One way to measure the decoherence suffered by two histories is through the
decoherence functional ($D$), which is \cite{Gell-Har, Har}

\begin{widetext}
\begin{equation*}
D(\alpha,\alpha')=\int_\alpha D\psi^1\int_{\alpha'}D\psi^2
\;\delta(\psi_f-\psi'_f)
e^{iS_0(\psi^1)}\rho_s(\psi_i,\psi'_i)e^{-iS_0(\psi^2)}\
\end{equation*}
\begin{equation}
\times\int d\xi_i \; d\xi'_i \int^{\xi^1}_{\xi_i}D\xi^1
\int^{\xi^2}_{\xi'_i}D\xi^2\;\delta(\xi^1-\xi^2)\;
e^{i[S_E(\xi^1)+S_I(\psi^1,\xi^1)]}\rho_E(\xi_i,\xi'_i)e^{-i[
S_E(\xi^2)+S_I(\psi^2,\xi^2)]}
\end{equation}
\end{widetext}

where $S_0$ is the free action of the system, $S_E$ is the action of environment, $S_I$ is the action for the system-environment interaction and $\rho_0$ and $\rho_E$ are the system and environment density matrices respectively. It is assumed that the system and environment are initially uncorrelated and therefore density matrix can be factorized.

The influence functional ($F$) is obtained through the integration of two final states of environment that are the same, ie $\xi^1=\xi^2=\xi$ \cite{CalHu,Fey-Hib}

\begin{equation*}
F(\psi^1,\psi^2)=e^{iS_{IF}}=\int d\xi \int d\xi_i \; d\xi'_i
\int^{\xi}_{\xi_i}D\xi^1\int^{\xi}_{\xi'_i}D\xi^2
\end{equation*}
\begin{equation}
 e^{i[S_E(\xi^1)+S_I(\psi^1,\xi^1)]}\rho_E(\xi_i,\xi'_i)e^{-i[S_E(\xi^2)+S_I(\psi^2,\xi^2)]}
\end{equation}  

Therefore, the decoherence functional is

\begin{equation*}
D(\alpha,\alpha')=\int_\alpha D\psi^1\int_{\alpha'}D\psi^2\;\delta(\psi_f-\psi'_f)
\end{equation*} 
\begin{equation}
 e^{iS_0(\psi^1)}\rho_s(\psi_i,\psi'_i)e^{-iS_0(\psi^2)}\;e^{iS_{IF}(\psi^1,\psi^2)}
\end{equation} 

The weak decoherence condition is recovered when

\begin{equation}
 e^{-Im[S_{IF}(\psi^1,\psi^2)]}<<1\;\;\;\Rightarrow\;\;\;Im[S_{IF}(\psi^1,
\psi^2)]>>1
\end{equation} 

If the interaction between system and environment can be written by a current coupling as

\begin{equation}
 S_I(\psi,\xi)=\int d^4x \;J(\psi(x))\xi(x)
\end{equation}

and the environment corresponds to free fields, then the influence functional can be written in terms of Jordan and Hadamard propagators as \cite{CalHu}

\begin{widetext}
\begin{equation*}
 S_{IF}(\psi^1,\psi^2)=\frac{i}{4}\int d^4xd^4x' \left[J(\psi^1)-J(\psi^2)
\right](x) \left[J(\psi^1)+J(\psi^2) \right](x') G(x,x')+
\end{equation*} 
\begin{equation}
+\frac{i}{4}\int d^4xd^4x'\left[J(\psi^1)-J(\psi^2)\right](x)\left[J(\psi^1)-J(\psi^2)\right](x')G_1(x,x')
\end{equation}
\end{widetext}

Since the currents $J(\psi)$ are real, all we need to consider to find the real part of the decoherence functional are propagators: the Jordan propagator ($G$) is imaginary while the Hadamard propagator ($G_1$) is real. Considering the factor $i$ before the influence functional, the imaginary part can be written as,

\begin{widetext}
\begin{equation}
Im(S_{IF})= \frac{1}{4}\int d^4x\;d^4x' \left[ J(\psi^1(x))-J(\psi^2(x))\right] 
\left[ J(\psi^1(x'))-J(\psi^2(x')) \right]  G_1(x,x') \label{fd}
\end{equation}
\end{widetext}

This is the expression to be computed to determine the decoherence of the scalar perturbations during inflation. It requires calculating the coupling current between the graviton and scalar fluctuations. This is the subject of the next Appendix.

\section{Inflation and cosmological perturbations}
\subsection{Quick review of Inflation}  

A period of accelerated expansion of early Universe, inflation, is proposed to resolve several inconsistencies existent in the standard Big Bang theory \cite{Bass, Guth}. The necessary condition achieve an accelerated expansion is $p=-\rho$. This condition yield the De-Sitter stage when the scale factor, which describes the evolution of the Universe, grows exponentially, $a\sim e^{Ht}$. 

This stage of evolution is dominated for the scalar field called inflaton
($\varphi$). Its energy density and pressure are given for 

\begin{equation}
\rho=\frac{1}{2}\dot{\varphi}^2+V(\varphi)\;\:\;;\;\;\;
p=\frac{1}{2}\dot{\varphi}^2-V(\varphi)
\end{equation} 

where $V(\varphi)$ is the potential energy of the inflaton.

In a expanding, homogeneous and isotropic space-time described for the plane FRW metric $ds^2=-dt^2+a^2(t)dx^2$, the inflaton follows the field equations,

\begin{equation}
 H^{2}= \frac{8 \pi}{3m^2_{pl}}\big[\frac{1}{2}\dot{\varphi}^{2}+V(\varphi)\big]\label{10}
\end{equation}
\begin{equation}
 \ddot{\varphi}+3H\dot{\varphi}+\frac{\partial V}{\partial \varphi}=0 \label{11}
\end{equation} 

where $H=\dot{a}/a$ is the Hubble factor (approximately constant during inflation) and $m_{pl}$ is the Planck mass. To reached the necessary amount of inflation the number of e-foldings, $N\equiv\log\frac{a_f}{a_i}$, must be $N > 60$.

The inflationary condition requires a sufficiently flat potential so that the potential energy dominates over the kinetic energy, $\dot{\varphi}^2<V(\varphi)$. This condition, known as slow-roll, is satisfied if

\begin{equation}
 \epsilon=\frac{m^2_{pl}}{16 \pi}\left( \frac{V_{,\varphi}}{V}\right)^2 <<1 \label{slowroll}
\end{equation}
\begin{equation}
\zeta=\frac{m^2_{pl}}{8 \pi}\frac{V_{,\varphi \varphi}}{V}<<1
\end{equation} 
 
where $\epsilon$ and $\zeta$ are the so-called slow-roll parameters. 

Using equation (\ref{slowroll}) to rewrite $V_{,\varphi}$ in terms of $\epsilon$ and neglecting the $\ddot{\varphi}$ term in (\ref{10}), the first slow roll parameter can be writer in terms of the kinetic and potential energies as

\begin{equation}
 \epsilon \approx \frac{\dot{\varphi}^2}{V} \label{epsilon}
\end{equation} 

Now, using that $V=m^2_{\varphi}\varphi^2$ in (\ref{slowroll}) the inflaton field results

\begin{equation}
 \varphi=m_{pl}/\sqrt{\epsilon} \label{inflaton}
\end{equation} 

and the Hubble factor is

\begin{equation}
 H\sim \frac{m_{\varphi}}{\sqrt{\epsilon}} \label{Hminflaton}
\end{equation}

It is convenient to define the conformal time, $dt=a d\eta$, as

\begin{equation}
-\frac{1}{aH}=\eta \label{conformaltime}
\end{equation} 

A derivative with respect to $\eta$ is denoted by $f'$. We also define $\mathcal{H}=a'/a=aH$.

Using the conformal time and the slow-roll parameters equation (\ref{11}) comes

\begin{equation}
 \varphi'\approx\sqrt{\epsilon}\frac{m_{pl}}{\eta} \label{inflatonderivative}
\end{equation}  

We will use those equations in next subsections and in the section four for the calculation of the decoherence functional.

\subsection{Invariant cosmological perturbations}

Perfectly homogeneous and isotropic space-time is only an idealization. This description can not explain the large structures observed in the Universe. One way to achieve a satisfactory explanation for the structure distribution is to include small perturbations in the FRW metric.

We will work with inhomogeneous small perturbations which are consistent with an inhomogeneous matter distribution and we will consider only lineal perturbations about the fields,

\begin{equation}
 \zeta=\zeta_0(t)+\delta\zeta(t,x)
\end{equation} 

The linear part of the perturbed FRW metric is \cite{MFB},

\begin{equation*}
ds^{2}=a^{2}(\eta)\left\lbrace
(1+2\phi)d\eta^{2}-2(S_i+B_{;i})dx^{i}d\eta \right.
\end{equation*} 
\begin{equation}
\left.-\left[(1-2\psi)\gamma_{ij}
+F_{i;j}+F_{j;i}+2E_{;ij}+h_{ij}\right]dx^{i}dx^{j}\right\rbrace \label{pertur}
\end{equation} 

where the $;$ sub index is the covariant derivative respect to the background space-time $\gamma_{ij}$ and $a(\eta)d\eta=dt$ is the conformal time. In the flat FRW space-time $\gamma_{ij}=\delta_{ij}$ and therefore the covariant derivative is the usual one.

The perturbations can be split into scalar, vector and tensor components according to their properties transformations in the spatial hyper surfaces. The scalar perturbations are $\phi$, $B$, $\psi$ and $E$. 

The vector component is given by $S$ and $F$ which satisfies $S^{;i}_i=F^{;i}_i=0$. The symmetric tensor $h_{ij}$ gives tensor perturbations with the constrains $h^i_i=0$ and $h^{;j}_{ij}=0$.

Those variables are gauge dependent. To describe the inhomogeneities of the universe through linear perturbations, we must first distinguish which of the quantities have a well defined physical interpretation, not related to a change of coordinates or a change in the system of reference. There is an infinite number of invariant quantities, but two commonly used are \cite{MFB}

\begin{equation}
 \Phi=\phi+\frac{1}{a}[(B-E')a]'\; \; ;\; \; \Psi=\psi-\frac{a'}{a}(B-E')
\end{equation} 

The reason for choosing these quantities is that in the \textit{Newtonian
gauge}, $B=E=0$, the two gauge invariant quantities coincide with the scalar perturbations in the metric that were not canceled, $\Phi=\phi$ and $\Psi=\psi$. Moreover, when the spatial part in the perturbation of the energy moment tensor is diagonal, the scalars perturbations $\phi$ and $\psi$ are equal and only one scalar degree of freedom in the metric remains. Furthermore,  a scalar quantity that is not included in the metric is already gauge invariant.

Regarding the tensor perturbations, they are gauge invariant for definition. Having zero trace and divergence, they do not have quantities that transform as scalars or vectors.

To get the scalar-tensor interaction we first must reduce the freedom degree of the scalar perturbations using the constraint that imposes the \textit{Newtonian gauge}. In this gauge, where $B=0$, all terms of superior order in $B$ will be canceled after we take its variation on the action, then is only necessary develop to linear order in terms of the perturbation $B$.

The translation of the ADM parameterization of the metric in terms of gauge invariant variables is as follows.
The shift function is

\begin{equation}
 N_i=a^2B_{,i}
\end{equation} 
the lapse function is
\begin{eqnarray}
N^2-N_iN^i&=&a^2\left(1+2\phi\right)\nonumber\\
N^2&=&a^2\left(1+2\phi+B_{,i}B_{,i}\right)\nonumber
\end{eqnarray} 
\begin{equation}
 N\approx a\left(1+\phi-\frac{1}{2}\phi^2\right)
\end{equation} 

and the extrinsic curvature tensor is

\begin{equation*}
K_{ij}=a\left\lbrace (1-\phi)B_{ij}- ^{(3)}\Gamma^k_{ij}B_{,k}  \left[\phi'\left(1-2\phi\right)\mathcal{H}-\right.\right.
\end{equation*}
\begin{equation*}
\left.\left.\phi\phi'+\phi\left(1-2\phi\right)\mathcal{H}
\frac{3}{2}\phi^2\mathcal{H}\right]\delta_{ij} \right.
\end{equation*}
\begin{equation}
\left.+\left(-1+\phi-\frac{3}{2}\phi^2\right)\mathcal{H}h_{ij}+\frac{1}{2}\left(-1+\phi-\frac{3}{4}\phi^2\right)h'_{ij}\right\rbrace 
\end{equation}

where 

\begin{equation}
 \Gamma^k_{ij}=\frac{1}{2}g^{kl}\left( g_{il,j}+g_{jl,i}-g_{ij,l} \right)
\end{equation}

is the spatial part of Christoffel's coefficients with 

\begin{equation}
 g_{ij}=-a^2(\eta)\left[(1-2\phi)\delta_{ij}+h_{ij}\right]
\end{equation}

the spatial part of the plane perturbed metric without vector perturbations.

In the \textit{Newtonian gauge} only one scalar perturbation in the metric remains. 
Now we need to find how this perturbation couples to perturbations on the inflaton field. 
To do this, it is necessary to give the perturbations in the energy-momentum tensor due to the perturbations in the inflaton field. 
Those are

\begin{equation}
\delta T^{0}_{0}=a^{-2}[-\varphi'^{2}_{0}\phi+ \varphi'_{0} \delta \varphi'
+V{,\varphi}a^{2}\delta \varphi]
\end{equation}
\begin{equation}
\delta T^{0}_{i}=a^{-2}\varphi'_{0}\delta \varphi_{,i} \label{nodiag2}
\end {equation}
\begin{equation}
\delta T^{i}_{j}=[\varphi'^{2}_{0}\phi- \varphi'_{0} \delta\varphi'+V_{,\varphi}a^{2}\delta \varphi]\delta^{i}_{j} \label{deltaTijSF}
\end{equation}

To find the dynamics of perturbations is need develop the Einstein's equations to linear order. The detail of this derivation is given in \cite{MFB}. The perturbed equations in the \textit{Newtonian gauge} are

\begin{equation}
 -3\mathcal{H}(\mathcal{H}\phi+\phi')+\nabla^{2}\phi=4 \pi m^{-2}_{pl} a^{2}\delta T^{0}_{0} \label{deltaT00}
\end{equation}
\begin{equation}
 (\mathcal{H}\phi+\phi')_{,i}=4 \pi m^{-2}_{pl} a^{2}\delta T^{0}_{i} \label{nodiag1}
\end{equation} 
\begin{equation}
\left[ \phi''+3\mathcal{H}\phi'+(2\mathcal{H}+\mathcal{H}^2)\phi\right] \delta^i_j=-4 \pi m^{-2}_{pl} a^{2} \delta T^{i}_{j} \label{deltaTij}
\end{equation} 

We will use those equations in the next subsection to find the dynamics of perturbations. 

\subsection{Scalar-Tensor interaction}

We left to appendix the details of derivations of the coupling current between the gauge invariant scalar mode $\phi$ and the gravitons. We start with the usual Einstein-Hilbert action written in ADM form \cite{MTW} plus the Klein-Gordon action for the inflaton

\begin{widetext}
\begin{equation*}
S=\frac{m^2_{pl}}{2}\int\;d^4x\left[
Ng^{1/2}(K^i_jK^j_i-K^2)+ \frac{1}{2}(g^{1/2}g^{ij}N)_{,i}(ln\;g)_{,j}N_{,i}(g^{1/2}g^{ij})_{,j} -\frac{1}{2}g^{1/2}N ^{(3)}\Gamma^k_{ij}g^{ij}_{,k}\right] 
\end{equation*} 
\begin{equation}
+\int\;d^4x \sqrt{-g}\left[ \frac{1}{2}g_{\mu\nu}
\varphi^{;\mu} \varphi^{;\nu} -V(\varphi) \right] 
\end{equation} 
\end{widetext}
where $2K_{ij}= N^{-1}[N_{i;j}+N_{j;i}-g'_{ij}]$ is the extrinsic curvature tensor, $N$ the lapse function and $N_{,i}$ the shift function.

The extrinsic curvature tensor provides no contributes to this coupling:
terms with scalar perturbations are accompanied with a Kronecker delta which when multiplied by tensor gravitons puts them terms of the form $\phi^2 h_{ii}$ and that due to the traceless of $h$ do not add to the coupling between them.

The other terms in the action contribute to the coupling.

\begin{equation*}
\frac{1}{2}m^2_{pl}N_{,i}(g^{1/2}g^{ij})_{,j}=\frac{1}{2}m^2_{pl}a^2\left[\phi_{,i}-\phi\phi_{,i}\right]\left\lbrace \left(1-3\phi+\frac{3}{2}\phi^2\right)\right.
\end{equation*}
\begin{equation*}
\left. \left[\left(1+2\phi+\phi^2\right)\delta^{ij}- \left(1+4\phi\right)h_{ij}\right]\right\rbrace _{,j}
\end{equation*} 

and retaining the terms of two scalar and one graviton is simplified in

\begin{equation}
\frac{1}{2}m^2_{pl}N_{,i}(g^{1/2}g^{ij})_{,j}=-\frac{1}{2}m^2_{pl}a^2\phi_{,i}\phi_{,j}h_{ij} \label{inter1}
\end{equation}

\begin{equation*}
-\frac{1}{2}m^2_{pl}\frac{1}{2}Ng^{1/2}\Gamma^k_{ij}g^{ij}_{,k}=-\frac{1}{4}m^2_{pl}a^2\left(1+\phi-\frac{1}{2}\phi^2\right)
\end{equation*} 
\begin{equation*}
\left(1-3\phi+\frac{3}{2}\phi^2\right) \left[\phi_{,k}\delta_{ij}-\phi_{,i}\delta_{kj}-\phi_{,j}\delta_{ki} \right.
\end{equation*}
\begin{equation*}
\left.+\frac{1}{2}\left(h_{ik,j}+h_{jk,i}-h_{ij,k}\right)\right]
\end{equation*} 
\begin{equation*}
\left[\left(2\phi_{,k}+8\phi\phi_{,k}\right)\delta_{ij}-4\phi_{,k}h_{ij}-\left(1+4\phi\right)h_{ij,k}\right]
\end{equation*} 

that keeping only the terms contributing to the interaction is reduced to

\begin{equation}
-\frac{1}{2}m^2_{pl}\frac{1}{2}Ng^{1/2}\Gamma^k_{ij}g^{ij}_{,k}=-2m^2_{pl}a^2\phi_{,i}\phi_{,j}h_{ij} \label{inter2}
\end{equation}

\begin{equation*}
\frac{1}{2}m^2_{pl}\frac{1}{2}(g^{1/2}g^{ij}N)_{,i}\frac{g_{,j}}{g}=\frac{1}{2}m^2_{pl}a^2\left\lbrace\left(1-3\phi+\frac{3}{2}\phi^2\right)\right.
\end{equation*} 
\begin{equation*}
\left.\left(1+\phi-\frac{1}{2}\phi^2\right)\left[\left(1+2\phi+4\phi^2\right)\delta_{ij}-\left(1+4\phi\right)h_{ij}\right]\right\rbrace_{,i}
 \end{equation*} 
\begin{equation*}
 \left(1+6\phi-12\phi^2\right)\left(-6\phi_{,j}+24\phi\phi_{,j}\right)
\end{equation*}
 
simplified in

\begin{equation}
\frac{1}{2}m^2_{pl}\frac{1}{2}(g^{1/2}g^{ij}N)_{,i}\frac{g_{,j}}{g}=3m^2_{pl}a^2\phi_{,i}\phi_{,j} \label{inter3}
\end{equation} 

Regarding the action of the inflaton, the only term that contributes to the interaction is  that contain $g^{ij}$, since none of the other components of the metric contains terms in $h_{ij}$. Then, for the inflaton

\begin{equation*}
\frac{1}{2}Ng^{1/2}g^{ij}\varphi_{,i}\varphi_{,j}=\frac{1}{2}a^2\left(1-3\phi\right)\left(1+\phi\right)\left(1+4\phi\right)\delta\varphi_{,i}\delta\varphi_{,j}h_{ij}
\end{equation*} 
\begin{equation}
\frac{1}{2}Ng^{1/2}g^{ij}\varphi_{,i}\varphi_{,j}=\frac{1}{2}a^2\delta\varphi_{,i}\delta\varphi_{,j}h_{ij} \label{inter4}
\end{equation} 

\subsection{Free scalar perturbations and CBR temperature}

The evolution of $\phi$ is obtained from the perturbed Einstein's equations. Subtract equation (\ref{deltaT00}) from (\ref{deltaTij}) and using (\ref{deltaTijSF}) for the spatial part of the perturbations of the energy-momentum tensor result on,

\begin{equation}
 \phi''+2\left( \frac{a}{\varphi'_0} \right)'\frac{\varphi'_0}{a}\phi'-\nabla^2
\phi + 2 \varphi'_0 \left( \frac{\mathcal{H}}{\varphi'_0} \right) '\phi=0
\end{equation} 

Making the change of variable $u=(a/\varphi'_0)\phi$, the last equation can be written as the Bessel equation with a time dependent mass term

\begin{equation}
 u''-\nabla^2 u - (z''/z)u=0; \; \; \; z=\mathcal{H}/a\varphi'_0
\end{equation} 

Expanding $u$ in its Fourier modes result the harmonic oscillator equations with a time depended frequency,

\begin{equation}
 u''_\textbf{k}+\omega^2_k (\eta)u_\textbf{k}=0, \;\;\; \omega_k
(\eta)=k^2-z''/z
\end{equation} 

The last equation can be resolved for two limits: one for short-wavelength modes (with $\lambda_{fis}=a/k<<H^{-1}$, ie $aH<<k$) and another for long wavelength modes when the modes are crossing the Hubble radius.

Using the slow-roll conditions to neglect the terms involving the time derivative of the inflaton we obtain

\begin{equation}
 \frac{z''}{z}\approx -2a^2H^2
\end{equation} 

and the equation for the modes $u_\textbf{k}$ results

\begin{equation}
 u''_\textbf{k}(\eta)=-(k^2+2a^2H^2)u_\textbf{k}(\eta)
\end{equation} 

In the short-wavelength limit the modes present an oscillating behavior 

\begin{equation}
 u_\textbf{k}(\eta)\simeq u_\textbf{k}(\eta_i)e^{ik(\eta-\eta_i)}
\end{equation} 

Then, in short wavelength limit the perturbation $\phi(\textbf{x},\eta)$ results

\begin{equation}
\phi(\textbf{x},\eta)=\int d^3k \frac{\varphi'}{a(\eta)} u_{\textbf{k}}(\eta_i)e^{ik(\eta-\eta_i)} e^{i\textbf{kx}}\hat{a}_k+c.c
\end{equation}

and using that during inflation $\varphi'/a(\eta)=\sqrt{\epsilon}m_{pl}$

\begin{equation}
\phi(\textbf{x},\eta)=\int d^3k\phi_{\textbf{k}}(\eta_i)e^{ik\eta}e^{i\textbf{kx}}\hat{a}_k+c.c
\label{freescalar}
\end{equation}

where

\begin{equation}
\phi_{\textbf{k}}(\eta_i)=H	u_{\textbf{k}}(\eta_i)
\end{equation}
Once the modes cross the horizon, $k_e\sim aH$, their amplitudes ($\phi_{\textbf{k}}(\eta_i)$) are frozen until they re-enter into the recombination era. At this stage their amplitudes are amplified and can be related to the inflationary stage through the equation \cite{Bard,MFB}

\begin{eqnarray}
\phi_{\textbf{k}}(\eta_i)&=&\frac{1+\frac{2}{3}\left[1+\omega(\eta_{rec})\right]^{-1}}{1+\frac{2}{3}\left[1+\omega(\eta_i)\right]^{-1}}\phi(\eta_{rec})\nonumber\\
\phi_{\textbf{k}}(\eta_i)&\approx&\frac{\dot{\varphi}^2}{V(\varphi)}\phi(\eta_{rec})
\end{eqnarray} 

where $\eta_i$ and $\eta_{rec}$ are the $k$ dependent times of initial and final
Hubble radius crossing and $\omega(\eta)=p/\rho$.

Moreover, using the Sachs-Wolfe effect \cite{Wei} we can relate the scalar perturbation with anisotropies in the Cosmic Background Radiation during recombination period as follows

\begin{eqnarray}
\phi(\eta_{rec})+\frac{\delta T}{T_0}&=&cte\nonumber\\
\frac{V(\varphi)}{\dot{\varphi}^2}\phi(\eta_i)+\frac{\delta T}{T_0}&=&cte \label{SW}
\end{eqnarray} 

\section{Decoherence functional computation}

After the rewrite the decoherence functional in terms of the coupling current given for (\ref{Sint}) and integrate in $d^3y$ and $d^3y'$ we obtain

\begin{widetext}
\begin{equation*}
Im(S_{IF})= \int  d\eta d\eta'd^3xd^3x' \frac{d^3k}{(2\pi)^3} \frac{d^3k'}{(2\pi)^3} \frac{d^3k''}{(2\pi)^3} \frac{d^3q}{(2\pi)^3}\frac{d^3p}{(2\pi)^3}
\frac{d^3q'}{(2\pi)^3}\frac{d^3p'}{(2\pi)^3} A^k_{ij}A^{k'}_{lm}q_ip_jq'_lp'_me^{i\textbf{k}\textbf{x}}e^{i\textbf{k}'\textbf{x}'} m^2_{pl}a^2(\eta)a^2(\eta')
\end{equation*}  
\begin{equation*}
\left\lbrace \alpha_{k''}\left[\delta(\textbf{k}''-\textbf{k})\delta(\textbf{k}''+\textbf{k}')+\delta(\textbf{k}''+\textbf{k})\delta(\textbf{k}''-\textbf{k}')\right]+\beta_{k''}\left[\delta(\textbf{k}''-\textbf{k})\delta(\textbf{k}''+\textbf{k}')-\delta(\textbf{k}''+\textbf{k})\delta(\textbf{k}''-\textbf{k}')\right] \right\rbrace 
\end{equation*} 
\begin{equation*}
\left(1+16m^2_{pl}\varphi'^{-2}_0\mathcal{H}^2-16m^2_{pl}\varphi'^{-2}_0qp+i32m^2_{pl}\varphi'^{-2}p\mathcal{H}\right)(\eta)
\left(1+16m^2_{pl}\varphi'^{-2}_0\mathcal{H}^2-16m^2_{pl}\varphi'^{-2}_0q'p'+i32m^2_{pl}\varphi'^{-2}p'\mathcal{H}\right)(\eta')
\end{equation*} 
\begin{equation}
\left\lbrace \phi^*_q(\eta) \phi^*_p(\eta) \phi_{q'}(\eta') \phi_{p'}(\eta') e^{i(\textbf{q}+\textbf{p})\textbf{x}}e^{-i(\textbf{q}'+\textbf{p}')\textbf{x}'}(\hat{a}^{-1}_q\hat{a}^{-1}_p-\hat{a}^{-2}_q\hat{a}^{-2}_p)(\hat{a}^{+1}_{q'}\hat{a}^{+1}_{p'}-\hat{a}^{+2}_{q'}\hat{a}^{+2}_{p'})+\text{other comb}\right\rbrace \label{fd1}
\end{equation} 
\end{widetext}

We use the Dirac's deltas to simplify the modes $\textbf{k}$ and $\textbf{k}'$ (only remains $\textbf{k}''$ and we recall it as $\textbf{k}$ to simplify). Later, we use the Fourier expansion on $\phi$ to integrate in $d^3x$ and $d^3x'$. Then, the decoherence functional results

\begin{widetext}
\begin{equation*}
Im(S_{IF})= \int  d\eta d\eta' \frac{d^3k}{(2\pi)^3} \frac{d^3q}{(2\pi)^3} \frac{d^3p}{(2\pi)^3}\frac{d^3q'}{(2\pi)^3}\frac{d^3p'}{(2\pi)^3}
 A^k_{ij}A^{k}_{lm} q_i p_j q'_l p'_m m^2_ {pl}a^2(\eta)a^2(\eta')
\end{equation*}
\begin{equation*}
\left(1+16m^2_{pl}\varphi'^{-2}_0\mathcal{H}^2-16m^2_{pl}\varphi'^{-2}_0qp+i32m^2_{pl}\varphi'^{-2}p\mathcal{H}\right)(\eta)
\left(1+16m^2_{pl}\varphi'^{-2}_0\mathcal{H}^2-16m^2_{pl}\varphi'^{-2}_0q'p'+i32m^2_{pl}\varphi'^{-2}p'\mathcal{H}\right)(\eta')
\end{equation*} 
\begin{equation*}
\left\lbrace (\alpha_k+\beta_k)\left[\phi^*_q(\eta) \phi^*_p(\eta) \phi_{q'}(\eta') \phi_{p'}(\eta')\delta(\textbf{k}+\textbf{q}+\textbf{p})\right. \left.\delta(-\textbf{k}-\textbf{q}'-\textbf{p}') (\hat{a}^{-1}_q\hat{a}^{-1}_p-\hat{a}^{-2}_q\hat{a}^{-2}_p) (\hat{a}^{+1}_{q'}\hat{a}^{+1}_{p'}-\hat{a}^{+2}_{q'}\hat{a}^{+2}_{p'})\right]\right. \end{equation*}
\begin{equation}
\left. +(\alpha_k-\beta_k)\left[ \phi^*_q(\eta) \phi^*_p(\eta) \phi_{q'}(\eta') \phi_{p'}(\eta')\delta(-\textbf{k}+\textbf{q}+\textbf{p})\right. \left.\delta(\textbf{k}-\textbf{q}'-\textbf{p}') (\hat{a}^{-1}_q\hat{a}^{-1}_p-\hat{a}^{-2}_q\hat{a}^{-2}_p) (\hat{a}^{+1}_{q'}\hat{a}^{+1}_{p'}-\hat{a}^{+2}_{q'}\hat{a}^{+2}_{p'})\right] +\text{other comb}\right\rbrace  \label{fd2}
\end{equation} 
\end{widetext}

Now the modes $\textbf{p}$ and $\textbf{p}'$ are simplified using deltas and the quantum operators are replaced by stochastic functions as in (\ref{stoc}).

\begin{widetext}
\begin{equation*}
Im(S_{IF})= \int  d\eta d\eta'\frac{d^3k}{(2\pi)^3}\frac{d^3q}{(2\pi)^3} \frac{d^3q'}{(2\pi)^3} A^k_{ij}A^{k}_{lm} m^2_ {pl}a^2(\eta)a^2(\eta')
\end{equation*}
\begin{equation*}
\left\lbrace (\alpha_k+\beta_k)q_i\arrowvert \textbf{k}+\textbf{q} \arrowvert_jq'_l\arrowvert \textbf{k}+\textbf{q}' \arrowvert_m \phi^*_q(\eta) \phi^*_{\arrowvert \textbf{k}+\textbf{q} \arrowvert}(\eta) \phi_{q'}(\eta') \phi_{\arrowvert \textbf{k}+\textbf{q}' \arrowvert}(\eta')\right.
\end{equation*} 
\begin{equation*}
\left.\left(1+16\frac{m^2_{pl}}{\varphi'^2_0}\mathcal{H}^2-16\frac{m^2_{pl}}{\varphi'^2_0}q\arrowvert \textbf{k}+\textbf{q} \arrowvert
+i32 \frac{m^2_{pl}}{\varphi'^2_0} \arrowvert \textbf{k} + \textbf{q} \arrowvert \mathcal{H}\right)(\eta)\left(1+16\frac{m^2_{pl}}{\varphi'^2_0}\mathcal{H}^2-16\frac{m^2_{pl}}{\varphi'^2_0}q'\arrowvert \textbf{k}+\textbf{q}' \arrowvert+i32\frac{m^2_{pl}}{\varphi'^2_0}\arrowvert \textbf{k}+\textbf{q}' \arrowvert\mathcal{H}\right)(\eta')\right.
\end{equation*} 
\begin{equation*}
L^3 \left. \left(f^*_{1q}e^{-i\theta_q}f^*_{1\arrowvert \textbf{k}+\textbf{q} \arrowvert}e^{-i\theta_{\arrowvert \textbf{k}+\textbf{q} \arrowvert}}-f^*_{2q}e^{-i\theta_q}f^*_{2\arrowvert \textbf{k}+\textbf{q} \arrowvert}e^{-i\theta_{\arrowvert \textbf{k}+\textbf{q} \arrowvert}}\right) 
L^3  \left(f_{1q'}e^{i\theta_{q'}}f_{1\arrowvert \textbf{k}+\textbf{q}' \arrowvert}e^{i\theta_{\arrowvert \textbf{k}+\textbf{q}' \arrowvert}}-f_{2q'}e^{i\theta_{q'}}f_{2\arrowvert \textbf{k}+\textbf{q}' \arrowvert}e^{i\theta_{\arrowvert \textbf{k}+\textbf{q}' \arrowvert}}\right)\right. 
\end{equation*} 
\begin{equation*}
\left.+(\alpha_k-\beta_k)q_i\arrowvert \textbf{k}-\textbf{q} \arrowvert_jq'_l\arrowvert \textbf{k}-\textbf{q}' \arrowvert_m \phi^*_q(\eta) \phi^*_{\arrowvert \textbf{k}-\textbf{q} \arrowvert}(\eta) \phi_{q'}(\eta') \phi_{\arrowvert \textbf{k}-\textbf{q}' \arrowvert}(\eta')\right.
\end{equation*} 
\begin{equation*}
\left.\left(1+16\frac{m^2_{pl}}{\varphi'^2_0}\mathcal{H}^2-16\frac{m^2_{pl}}{\varphi'^2_0}q\arrowvert \textbf{k}-\textbf{q} \arrowvert
+i32\frac{m^2_{pl}}{\varphi'^2_0}\arrowvert \textbf{k}-\textbf{q} \arrowvert\mathcal{H}\right)(\eta)\left(1+16\frac{m^2_{pl}}{\varphi'^2_0}\mathcal{H}^2-16\frac{m^2_{pl}}{\varphi'^2_0}q'\arrowvert \textbf{k}-\textbf{q}' \arrowvert+i32\frac{m^2_{pl}}{\varphi'^2_0}\arrowvert \textbf{k}-\textbf{q}' \arrowvert\mathcal{H}\right)(\eta')\right.
\end{equation*} 
\begin{equation}
L^3 \left. \left(f^*_{1q}e^{-i\theta_q}f^*_{1\arrowvert \textbf{k}-\textbf{q} \arrowvert}e^{-i\theta_{\arrowvert \textbf{k}-\textbf{q} \arrowvert}}-f^*_{2q}e^{-i\theta_q}f^*_{2\arrowvert \textbf{k}-\textbf{q} \arrowvert}e^{-i\theta_{\arrowvert \textbf{k}-\textbf{q} \arrowvert}}\right)
L^3 \left(f_{1q'}e^{i\theta_{q'}}f_{1\arrowvert \textbf{k}-\textbf{q}' \arrowvert}e^{i\theta_{\arrowvert \textbf{k}-\textbf{q}' \arrowvert}}-f_{2q'}e^{i\theta_{q'}}f_{2\arrowvert \textbf{k}-\textbf{q}' \arrowvert}e^{i\theta_{\arrowvert \textbf{k}-\textbf{q}' \arrowvert}}\right) \right\rbrace \label{fd3}
\end{equation} 
\end{widetext}

where $L^3$ is the volume where the stochastic functions were defined. 

Using equation (\ref{CBR}) to rewrite the decoherence functional in terms of the CBR temperature fluctuations we obtain

\begin{widetext}
\begin{equation*}
Im(S_{IF}) \approx m^2_{pl}  \left[\frac{\dot{\varphi}^2}{V}\right]^4 L^3 \int d^3k \;d^3q \; d\eta \; d\eta'  a^2(\eta) a^2(\eta') .
\end{equation*} 
\begin{equation*}
\left\lbrace \left(\alpha_k + \beta_k\right)\left[A^k_{ij}q_i\arrowvert \textbf{k}+\textbf{q} \arrowvert_j\right]^2 e^{iq(\eta'-\eta)}e^{i\arrowvert \textbf{k}+\textbf{q}\arrowvert(\eta'-\eta)}\left[ \frac{\delta T_1(q)}{T_0}\frac{\delta T_1(\arrowvert\textbf{k}+\textbf{q}\arrowvert)}{T_0}-\frac{\delta T_2(q)}{T_0}\frac{\delta T_2(\arrowvert\textbf{k}+\textbf{q}\arrowvert)}{T_0}\right]^2 \right.
\end{equation*}
\begin{equation*}
\left.\left(1+16\frac{m^2_{pl}}{\varphi'^2_0}\mathcal{H}^2-16\frac{m^2_{pl}}{\varphi'^2_0}q\arrowvert \textbf{k}+\textbf{q} \arrowvert
+i32 \frac{m^2_{pl}}{\varphi'^2_0} \arrowvert \textbf{k} + \textbf{q} \arrowvert \mathcal{H}\right)(\eta)\left(1+16\frac{m^2_{pl}}{\varphi'^2_0}\mathcal{H}^2-16\frac{m^2_{pl}}{\varphi'^2_0}q'\arrowvert \textbf{k}+\textbf{q}' \arrowvert+i32\frac{m^2_{pl}}{\varphi'^2_0}\arrowvert \textbf{k}+\textbf{q}' \arrowvert\mathcal{H}\right)(\eta')\right.
\end{equation*}
\begin{equation*}
\left. +\left(\alpha_k-\beta_k\right) \left[ A^k_{ij}q_i \arrowvert \textbf{k}- \textbf{q} \arrowvert_j \right]^2 e^{iq(\eta'-\eta)}e^{i\arrowvert \textbf{k}-\textbf{q}\arrowvert(\eta'-\eta)}\left[ \frac{\delta T_1(q)}{T_0}\frac{\delta T_1(\arrowvert\textbf{k}-\textbf{q}\arrowvert)}{T_0}-\frac{\delta T_2(q)}{T_0}\frac{\delta T_2(\arrowvert\textbf{k}-\textbf{q}\arrowvert)}{T_0}\right]^2\right.
\end{equation*}
\begin{equation}
\left(1+16\frac{m^2_{pl}}{\varphi'^2_0}\mathcal{H}^2-16\frac{m^2_{pl}}{\varphi'^2_0}q\arrowvert \textbf{k}-\textbf{q} \arrowvert
+ i 32 \frac{m^2_{pl}}{\varphi'^2_0} \arrowvert \textbf{k} - \textbf{q} \arrowvert \mathcal{H}\right)(\eta)\left.\left(1+16\frac{m^2_{pl}}{\varphi'^2_0}\mathcal{H}^2-16\frac{m^2_{pl}}{\varphi'^2_0}q'\arrowvert \textbf{k}-\textbf{q}' \arrowvert + i32 \frac{m^2_{pl}}{\varphi'^2_0}\arrowvert \textbf{k}-\textbf{q}' \arrowvert\mathcal{H}\right)(\eta')\right\rbrace \label{fd4}
\end{equation} 
\end{widetext}

The histories of the CBR fluctuations are taken as follows: we consider that this histories are the same except in a small window centered at $\textbf{q}_0$. Moreover, one of this histories is the observed Harrison-Zel'dovich spectrum and the another history remains unspecified. 

The limits in the integral are defined by the modes which stay inside the Hubble horizon during inflation. The long wavelength modes cross the horizon at the beginning of inflation, therefore the minimum mode is the order of the horizon at the initials times of inflation, $k_{min}\sim a_iH$; by other hand, the short wavelength modes exit at the end of inflation and give as the maximum possible wave number $k_{Max}\sim a_fH$. Then, the integral in $d^3k$ takes the limits $\left[a_iH;a_fH\right]$.

\begin{widetext}
\begin{equation*}
Im(S_{IF}) \approx \frac{m^2_{pl}\epsilon^4}{H^4} L^3 (\delta q_0)^3 \left[\frac{\delta T_1(q_0)}{T_0}-\frac{\delta T_2(q_0)}{T_0}\right]^2\int d^3k \; d\eta \; d\eta' \frac{1}{\eta^2}\frac{1}{\eta'^2}
\end{equation*} 
\begin{equation*}
\left\lbrace \frac{1+k^2\eta\eta'+k(\eta-\eta')}{k^3\eta\eta'}\left[A^k_{ij}q_i\arrowvert \textbf{k}+\textbf{q} \arrowvert_j\right]^2 e^{iq_0(\eta'-\eta)}e^{i\arrowvert \textbf{k}+\textbf{q}_0\arrowvert(\eta'-\eta)}\frac{1}{\arrowvert \textbf{k}+\textbf{q}_0\arrowvert^3} \right.
\end{equation*}
\begin{equation*}
\left.\left(1+\frac{16}{\epsilon}-\frac{16}{\epsilon}\eta^2q_0\arrowvert \textbf{k}+\textbf{q}_0 \arrowvert-\frac{i32}{\epsilon} \eta \arrowvert \textbf{k} + \textbf{q}_0 \arrowvert \right) \left(1+\frac{16}{\epsilon}-\frac{16}{\epsilon}\eta'^2q_0\arrowvert \textbf{k}+\textbf{q}_0 \arrowvert-\frac{i32}{\epsilon} \eta' \arrowvert \textbf{k} + \textbf{q}_0 \arrowvert \right)\right.
\end{equation*}
\begin{equation*}
\left. +\frac{1+k^2\eta\eta'-k(\eta-\eta')}{k^3\eta\eta'} \left[ A^k_{ij}q_i \arrowvert \textbf{k}- \textbf{q} \arrowvert_j \right]^2 e^{iq_0(\eta'-\eta)}e^{i\arrowvert \textbf{k}-\textbf{q}_0\arrowvert(\eta'-\eta)}\frac{1}{\arrowvert \textbf{k}-\textbf{q}_0\arrowvert^3}\right.
\end{equation*}
\begin{equation}
\left.\left(1+\frac{16}{\epsilon}-\frac{16}{\epsilon}\eta^2q_0\arrowvert \textbf{k}-\textbf{q}_0 \arrowvert-\frac{i32}{\epsilon} \eta \arrowvert \textbf{k} - \textbf{q}_0 \arrowvert \right) \left(1+\frac{16}{\epsilon}-\frac{16}{\epsilon}\eta'^2q_0\arrowvert \textbf{k}-\textbf{q}_0 \arrowvert-\frac{i32}{\epsilon} \eta' \arrowvert \textbf{k}-\textbf{q}_0 \arrowvert \right)\right\rbrace \label{fd5}
\end{equation} 
\end{widetext}

where we use  equation (\ref{epsilon}) for $\dot{\varphi}^2/V$, equation (\ref{conformaltime}) for the scale factor, equation (\ref{inflatonderivative}) for $\varphi'$ and we replace $\alpha\pm\beta$ for equation (\ref{alfa}).

At this point there only remains to integrate over the conformal times and in the $\textbf{k}$ mode. First it is convenient to integrate over the conformal times. This integral must be done from  $\eta_i$ until each mode crosses the horizon, $\eta_e(k)\sim-1/k$. Since at the beginning of inflation the conformal time is extremely large in absolute value, we may keep only the quadratic term in the coupling current.

\begin{widetext}
\begin{equation*}
Im(S_{IF}) \approx \frac{m^2_{pl}\epsilon^4}{H^4} L^3 (\delta q_0)^3 \left[\frac{\delta T_1(q_0)}{T_0}-\frac{\delta T_2(q_0)}{T_0}\right]^2\int d^3k \; \int^{\eta_e}_{\eta_i}\int^{\eta'_e}_{\eta_i}d\eta \; d\eta' \frac{1}{\eta^2}\frac{1}{\eta'^2}
\end{equation*} 
\begin{equation*}
\left\lbrace \frac{1+k^2\eta\eta'+k(\eta-\eta')}{k^3\eta\eta'}\left[A^k_{ij}q_i\arrowvert \textbf{k}+\textbf{q} \arrowvert_j\right]^2 e^{iq_0(\eta'-\eta)}e^{i\arrowvert \textbf{k}+\textbf{q}_0\arrowvert(\eta'-\eta)}\frac{1}{\arrowvert \textbf{k}+\textbf{q}_0\arrowvert^3} \frac{16}{\epsilon^2}q^2_0\arrowvert \textbf{k}+\textbf{q}_0\arrowvert^2 \eta^2\eta'^2 \right.
\end{equation*}
\begin{equation}
\left. +\frac{1+k^2\eta\eta'-k(\eta-\eta')}{k^3\eta\eta'} \left[ A^k_{ij}q_i \arrowvert \textbf{k}- \textbf{q} \arrowvert_j \right]^2 e^{iq_0(\eta'-\eta)}e^{i\arrowvert \textbf{k}-\textbf{q}_0\arrowvert(\eta'-\eta)}\frac{1}{\arrowvert \textbf{k}-\textbf{q}_0\arrowvert^3}\frac{16}{\epsilon^2}q^2_0\arrowvert \textbf{k}-\textbf{q}_0\arrowvert^2 \eta^2\eta'^2\right\rbrace
\end{equation}
\end{widetext}

where $\eta_e\sim-1/k$ and $\eta'_e\sim-1/\arrowvert\textbf{k}\pm\textbf{q}_0\arrowvert$ are the times when the modes $k$ and $\arrowvert\textbf{k}\pm\textbf{q}_0\arrowvert$ crosses the horizon.

After integrating and taking the real part, the decoherence functional is

\begin{equation}
Im(S_{IF}) \approx \frac{m^2_{pl}\epsilon^4}{H^4} L^3 (\delta q_0)^3 I(\eta_i,q_0)\;\left[\frac{\delta T_1(q_0)}{T_0}-\frac{\delta T_2(q_0)}{T_0}\right]^2 
\end{equation}

where  $I(\eta_i,q_0)$  is 

\begin{widetext}
\begin{equation*}
I(\eta_i,q_0)=\int d^3k \left\lbrace\frac{16q^2_0}{\epsilon^2k^3\arrowvert\textbf{k}+\textbf{q}_0\arrowvert} \left[A^k_{ij}q_i\arrowvert \textbf{k}+\textbf{q} \arrowvert_j\right]^2\right.
\end{equation*} 
\begin{equation*} 
\left.
\left[\left(\mathrm{Ci}[1+\frac{\arrowvert\textbf{k}+\textbf{q}_0\arrowvert}{q_0}]-\mathrm{Ci}[q_0\eta_i(1+\frac{\arrowvert\textbf{k}+\textbf{q}_0\arrowvert}{q_0})]\right)\left(\mathrm{Ci}[1+\frac{q_0}{\arrowvert\textbf{k}+\textbf{q}_0\arrowvert}]-\mathrm{Ci}[q_0\eta_i(1+\frac{\arrowvert\textbf{k}+\textbf{q}_0\arrowvert}{q_0})]\right)\right.\right.
\end{equation*}
\begin{equation*}
\left.+\frac{k^2}{\left(q_0+\arrowvert\textbf{k}+\textbf{q}_0\arrowvert\right)^2} \left(\cos[\frac{\arrowvert \textbf{k}+\textbf{q}_0\arrowvert^2-q^2_0}{q^2_0\arrowvert \textbf{k}+\textbf{q}_0\arrowvert}]-\cos[(q_0+\arrowvert \textbf{k}+\textbf{q}_0\arrowvert)(\eta_i+\frac{1}{q_0})]-\cos[(q_0+\arrowvert \textbf{k}+\textbf{q}_0\arrowvert)(\eta_i+\frac{1}{\arrowvert \textbf{k}+\textbf{q}_0\arrowvert})]+1\right)\right.
\end{equation*}
\begin{equation*}
\left.+\frac{k}{q_0+\arrowvert\textbf{k}+\textbf{q}_0\arrowvert}\left(\mathrm{Ci}[1+\frac{\arrowvert\textbf{k}+\textbf{q}_0\arrowvert}{q_0}]-\mathrm{Ci}[q_0\eta_i(1+\frac{\arrowvert \textbf{k}+\textbf{q}_0\arrowvert}{q_0})]\right)\left(-\sin[1+\frac{q_0}{\arrowvert \textbf{k}+\textbf{q}_0\arrowvert}]-\sin[q_0\eta_i(1+\frac{\arrowvert \textbf{k}+\textbf{q}_0\arrowvert}{q_0})]\right)\right. 
\end{equation*}
\begin{equation*}
\left.\left.+\frac{k}{q_0+\arrowvert\textbf{k}+\textbf{q}_0\arrowvert}\left(\mathrm{Ci}[1+\frac{q_0}{\arrowvert\textbf{k}+\textbf{q}_0\arrowvert}]-\mathrm{Ci}[q_0\eta_i(1+\frac{\vert\textbf{k}+\textbf{q}_0\vert}{q_0})]\right)\left(\sin[1+\frac{\arrowvert \textbf{k}+\textbf{q}_0\arrowvert}{q_0}] +\sin[q_0\eta_i(1+\frac{\arrowvert \textbf{k}+\textbf{q}_0\arrowvert}{q_0})]\right)\right]\right.
\end{equation*} 
\begin{equation*}
\left.+\frac{16q^2_0}{\epsilon^2k^3\arrowvert\textbf{k}-\textbf{q}_0\arrowvert} \left[A^k_{ij}q_i\arrowvert \textbf{k}-\textbf{q} \arrowvert_j\right]^2 \left[\left(\mathrm{Ci}[1+\frac{\arrowvert\textbf{k}-\textbf{q}_0\arrowvert}{q_0}]-\mathrm{Ci}[q_0\eta_i(1+\frac{\arrowvert\textbf{k}-\textbf{q}_0\arrowvert}{q_0})]\right)\left(\mathrm{Ci}[1+\frac{q_0}{\arrowvert\textbf{k}-\textbf{q}_0\arrowvert}]-\mathrm{Ci}[q_0\eta_i(1+\frac{\arrowvert\textbf{k}-\textbf{q}_0\arrowvert}{q_o})]\right)\right.\right.
\end{equation*}
\begin{equation*}
\left.+\frac{k^2}{\left(q_0+\arrowvert \textbf{k}-\textbf{q}_0\arrowvert\right)^2} \left(\cos[\frac{\arrowvert \textbf{k}-\textbf{q}_0\arrowvert^2-q^2_0}{q^2_0\arrowvert \textbf{k}-\textbf{q}_0\arrowvert}]-\cos[(q_0+\arrowvert \textbf{k}-\textbf{q}_0\arrowvert)(\eta_i+\frac{1}{q_0})]-\cos[(q_0+\arrowvert \textbf{k}-\textbf{q}_0\arrowvert)(\eta_i+\frac{1}{\arrowvert \textbf{k}-\textbf{q}_0\arrowvert})]+1\right)\right.
\end{equation*}
\begin{equation*}
\left.+\frac{k}{q_0+\arrowvert\textbf{k}-\textbf{q}_0\arrowvert}\left(\mathrm{Ci}[1+\frac{\arrowvert\textbf{k}-\textbf{q}_0\arrowvert}{q_0}]-\mathrm{Ci}[q_0\eta_i(1+\frac{\arrowvert \textbf{k}-\textbf{q}_0\arrowvert}{q_0})]\right)\left(-\sin[1+\frac{q_0}{\arrowvert \textbf{k}-\textbf{q}_0\arrowvert}]-\sin[q_0\eta_i(1+\frac{\arrowvert \textbf{k}-\textbf{q}_0\arrowvert}{q_0})]\right)\right. 
\end{equation*}
\begin{equation}
\left.\left.+\frac{k}{q_0+\arrowvert\textbf{k}-\textbf{q}_0\arrowvert}\left(\mathrm{Ci}[1+\frac{q_0}{\arrowvert\textbf{k}-\textbf{q}_0\arrowvert}]-\mathrm{Ci}[q\eta_i(1+\frac{\vert\textbf{k}-\textbf{q}_0\vert}{q_0})]\right)\left(\sin[1+\frac{\arrowvert \textbf{k}-\textbf{q}_0\arrowvert}{q_0}]+ \sin[q_0\eta_i(1+\frac{\arrowvert \textbf{k}-\textbf{q}_0\arrowvert}{q_0})]\right)\right]\right\rbrace \label{fd6}
\end{equation} 
\end{widetext}

where $\mathrm{Ci}(x)=-\int^\infty_x \frac{\cos t}{t}dt$ is the integral cosine which is the real part of the exponential integral $\mathrm{Ei}(\pm ix)=\mathrm{Ci}(x)\pm i\mathrm{Si}(x)$\cite{handbook}. This function oscillates and has a bounded asymptotic behavior, then all terms that contain this function are subdominant. The terms that comes of the product between the exponentials have a oscillating behavior too, but there is one term that no oscillates. It comes from when $\eta=\eta'=\eta_i$, so we retain only this term. The parametric dependence on $\eta_i$ is lost because it only appears in the integral cosines and sines.

Putting back the \textit{other comb} in the calculation and having into account that $A^k_{ij}q_i\arrowvert \textbf{k}-\textbf{q} \arrowvert_j = \vert\textbf{k} \times\textbf{q} \vert$, there only remains to compute

\begin{widetext}
\begin{equation*}
I(q_0)=\int d^3k\; \frac{\vert\textbf{k}\times\textbf{q}_0\vert^2q^2_0}{\epsilon^2k^3}\left[\frac{1}{\vert\textbf{k}+\textbf{q}_0\vert}\frac{k^2}{(q_0+\vert\textbf{k}+\textbf{q}_0\vert)^2}+\frac{1}{\vert\textbf{k}+\textbf{q}_0\vert}\frac{k^2}{(q_0-\vert\textbf{k}+\textbf{q}_0\vert)^2}\right.
\end{equation*}
\begin{equation}
\left.+\frac{1}{\vert\textbf{k}-\textbf{q}_0\vert}\frac{k^2}{(q_0+\vert\textbf{k}-\textbf{q}_0\vert)^2}+\frac{1}{\vert\textbf{k}-\textbf{q}_0\vert}\frac{k^2}{(q_0-\vert\textbf{k}-\textbf{q}_0\vert)^2}\right]
\label{fd7}
\end{equation}
\end{widetext}

\end{document}